\documentclass{PoS}

\usepackage{url}

\title{Structure of Exotic Nuclei: A Theoretical Review}

\ShortTitle{Structure of Exotic Nuclei: A Theoretical Review}

\author{
	Shan-Gui Zhou\thanks{Supported by the NSF of China (11275248, 
                             11525524, 11621131001 and 11647601), the 973 
                             Program of China (2013CB834400), 
                             the Key Research Program of Frontier Sciences of CAS,
                             the HPC Cluster 
                             of SKLTP/ITP-CAS and the Supercomputing Center, 
                             CNIC of CAS.} \\
        \hangindent 1.0em
        CAS Key Laboratory of Frontiers in Theoretical Physics,
         Institute of Theoretical Physics, Chinese Academy of Sciences, 
	 Beijing 100190, China \\
        School of Physical Sciences, University of Chinese Academy of Sciences, 
	 Beijing 100049, China \\
        Center of Theoretical Nuclear Physics, National Laboratory
         of Heavy Ion Accelerator, Lanzhou 730000, China \\
        Synergetic Innovation Center for Quantum Effects and Application,
         Hunan Normal University, Changsha, 410081, China \\\\
        E-mail: \email{sgzhou@itp.ac.cn}}

\abstract{
The study of exotic nuclei---nuclei with the ratio of
neutron number $N$ to proton number $Z$ deviating much from that of
those found in nature---is at the forefront of nuclear physics research
because it can not only reveal novel nuclear properties and thus enrich
our knowledge of atomic nuclei, but also help us to understand the origin
of chemical elements in the nucleosynthesis.
With the development of radioactive ion beam facilities around the world, 
more and more unstable nuclei become experimentally accessible.
Many exotic nuclear phenomena have been observed or predicted in nuclei
far from the $\beta$-stability line, such as neutron or proton halos,
the shell evolution and changes of nuclear magic numbers,
the island of inversion, soft-dipole excitations, clustering effects,
new radioactivities, giant neutron halos, the shape decoupling between core 
and valence nucleons in deformed halo nuclei, etc.
In this contribution, I will present a review of theoretical study of
exotic nuclear structure. I will first introduce characteristic features
and new physics connected with exotic nuclear phenomena:
the weakly-bound feature, the large-spatial extension in halo nuclei,
deformation effects in halo nuclei, the shell evolution, new radioactivities 
and clustering effects. Then I will highlight some recent progresses 
corresponding to these features.
}

\FullConference{The 26th International Nuclear Physics Conference\\
		11-16 September, 2016\\
		Adelaide, Australia}

\begin{document}

\section{Introduction}

\begin{figure}
\begin{center}
\includegraphics[width=0.70\textwidth]{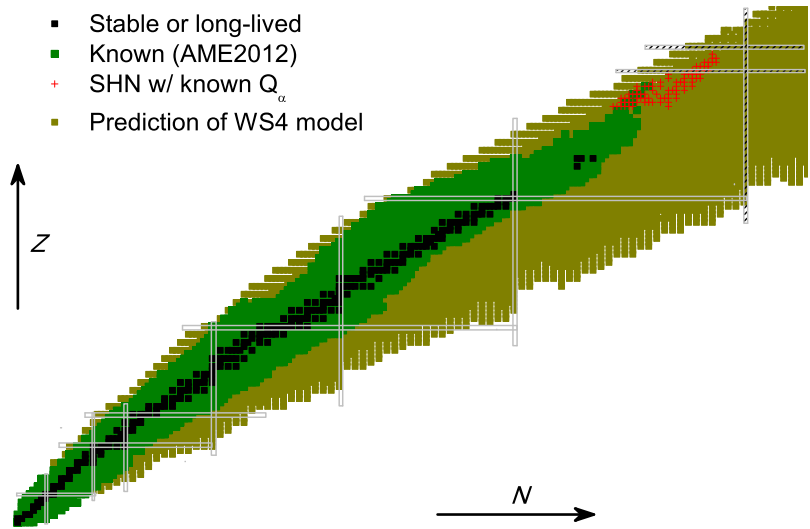}
\end{center}
\caption{(Color online)
Nuclear chart consisting of stable or long-lived nuclides, known and 
unstable ones produced in laboratories [including superheavy nuclei (SHN)]
\cite{Audi2012_ChinPhysC36-1157,Audi2012_ChinPhysC36-1287,
Wang2012_ChinPhysC36-1603} 
and those predicted by the Weiszacker-Skyrme (WS4) mass model 
\cite{Wang2014_PLB734-215}. Courtesy of Ning Wang.}
\label{fig:WS4}
\end{figure}

In nature, there are less than three hundred stable or long-lived nuclides 
which are along the valley of stability in the nuclear chart.
When those unstable nuclei, 
the number is now close to three thousand
\cite{Thoennessen2013_RPP76-056301},
are explored, many exotic nuclear phenomena have been observed. 
The most famous exotic nucleus is $^{11}$Li in which the halo structure was 
identified \cite{Tanihata1985_PRL55-2676}. 
Theoretically, many more nuclei are predicted to be bound. 
In Fig.~\ref{fig:WS4} is shown the prediction from the Weiszacker-Skyrme (WS4) 
mass model \cite{Wang2014_PLB734-215} which is one of the best nuclear mass 
models on the market. 
Many other models, e.g., non-relativistic \cite{Erler2012_Nature486-509}
and relativistic density functional theories \cite{
Afanasjev2013_PLB726-680,Qu2013_SciChinaPMA56-2031,
Lu2015_PRC91-027304} 
have also been used to explore the border of nuclear chart.
Nowadays, the study of the properties of these exotic nuclei is at the forefront
of nuclear physics research because it can not only reveal new physics 
but also lead to new insights on the nucleosynthesis.

In this contribution, I will first discuss the physics connected with exotic
nuclear phenomena in Section~\ref{sec:physics}.  
According to my personal point of view, six features will be illustrated
concerning exotic nuclear structure.
Then I'll highlight some recent progresses corresponding to each of these
features in Section~\ref{sec:highlights}. 
Finally I will discuss perspectives in Section~\ref{sec:summary}.

\section{\label{sec:physics}Physics in exotic nuclear structure}

\begin{figure}
\begin{center}
\includegraphics[width=1.00\textwidth]{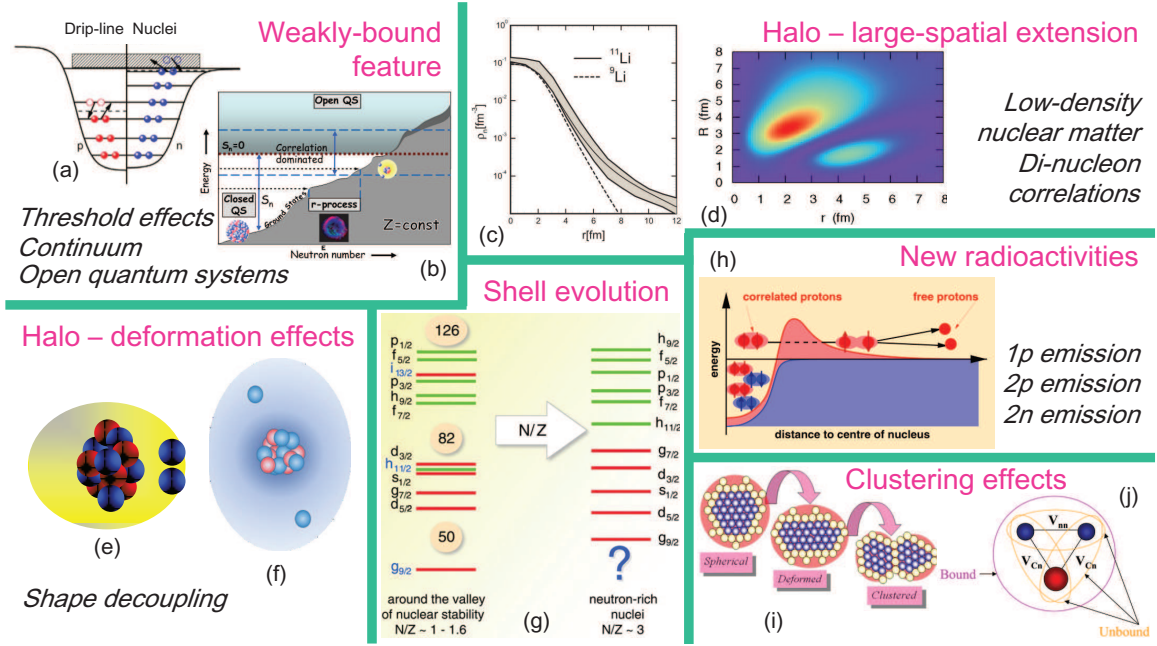}
\end{center}
\caption{(Color online)
Physics in exotic nuclear structure. 
(a), (b), (c), (d), (g), (h), (i) and (j) are 
taken from Refs.~\cite{Meng2006_PPNP57-470,Michel2009_JPG36-013101,
Meng1996_PRL77-3963,Hagino2007_PRL99-022506,Dobaczewski2007_PPNP59-432,
Blank2008_RPP71-046301,Freer2007_RPP70-2149,Johnson2004_PR389-1}, respectively.
(f): Courtesy of Junchen Pei.
} 
\label{fig:physics}
\end{figure}

The first important characteristic of exotic nuclei is certainly 
\emph{the weakly-bound feature}. 
In unstable nuclei, particularly in those close to drip lines, e.g., 
to the neutron drip line, the neutron Fermi surface is very close to the threshold, 
as seen in Fig.~\ref{fig:physics}(a) 
\cite{Meng2006_PPNP57-470,Meng2015_JPG42-093101}.
Therefore the contribution from continua becomes more and more important. 
In Fig.~\ref{fig:physics}(b), 
the neutron separation energy $S_n$, equivalent to the neutron Fermi energy, 
is shown schematically as a function of neutron number 
\cite{Dobaczewski2007_PPNP59-432,Michel2009_JPG36-013101}. 
Larger neutron excess results in smaller $S_n$, i.e., the valence neutron(s) 
is (are) more easily knocked out and the nucleus is more easily coupled to the 
scattering environment, thus making exotic nuclei open quantum systems which are
very much involved in the studies of nuclear reactions and nucleosynthesis.

Halo nuclei are characterized by a \emph{large spatial extension}, see 
Fig.~\ref{fig:physics}(c) for an example for $^{11}$Li \cite{Meng1996_PRL77-3963}. 
In neutron halo nuclei, there appears pure neutron matter with a very low density,
surrounding a dense core 
\cite{Meng2006_PPNP57-470,Meng2015_JPG42-093101,
Meng1996_PRL77-3963,Vretenar2005_PR409-101}. 
Similar to what happens in low density infinite nuclear matter or neutron matter
\cite{Sun2010_PLB683-134,Matsuo2006_PRC73-044309,Sun2012_PRC86-014305}, 
pair condensate or strong di-neutron correlations may occur in finite nuclei 
with halo structure [Fig.~\ref{fig:physics}(d)] 
\cite{Hagino2007_PRL99-022506,Sagawa2015_EPJA51-102}. 
In addition, the oscillation between the core and the low density neutron matter
leads to some soft dipole modes, also known as pygmy dipole resonances, which 
have been discussed a lot in INPC2016. Experimentally these features 
have been explored in, e.g., Refs.~\cite{Nakamura2006_PRL96-252502,
Kanungo2015_PRL114-192502}. 

Most known nuclei are deformed \cite{Zhou2016_PS91-063008}. 
What kind of new features can deformation effects bring to exotic nuclei,
in particular, to halo nuclei? 
Note that in recent years, more candidates of deformed halo nuclei have been 
identified; examples are 
$^{31}$Ne \cite{Nakamura2014_PRL112-142501} and 
$^{37}$Mg \cite{Kobayashi2014_PRL112-242501}. 
For \emph{deformation effects in halo nuclei}, I will focus on theoretical 
predictions on the shape decoupling [Fig.~\ref{fig:physics}(e) \& (f)]
\cite{Zhou2010_PRC82-011301R,Li2012_PRC85-024312,
Pei2013_PRC87-051302R,Pei2014_PRC90-024317}.

Shell structure is very important in the study of atomic nuclei which is 
characterized by large spin-orbit couplings. The spin-orbit couplings are 
closely connected with the nuclear surface diffuseness. 
It is therefore very natural that the spin-orbit splitting would change 
when going from the $\beta$-stability line to the drip lines 
because the nuclear surface could be more diffuse [Fig.~\ref{fig:physics}(g)] 
\cite{Dobaczewski2007_PPNP59-432}. This results in the \emph{shell evolution} 
in exotic nuclei and changes of nuclear magicity which can be hinted from,
e.g., separation energies \cite{Ozawa2000_PRL84-5493}. 
Certainly the shell evolution is also the result of many other important 
factors, like the tensor force 
\cite{
Peru2000_EPJA9-35,
Otsuka2005_PRL95-232502,
Colo2007_PLB646-227,
Sorlin2008_PPNP61-602,
Sagawa2014_PPNP76-76}.
It should be emphasized that shape evolution and shape coexistence are also 
related physical topics of exotic nuclei, which were discussed in a dedicated 
session in INPC2016.

Beyond the drip lines, nuclei are unbound with respect to nucleon(s) emission. 
This feature implies some \emph{new radioactivities} of which mostly discussed 
are one- or two-proton radioactivities, thanks to the Coulomb interaction which
leads to a Coulomb barrier hindering the escape of proton(s) from the parent 
nucleus [Fig.~\ref{fig:physics}(h)] 
\cite{Blank2008_RPP71-046301,Woods1997_ARNPS47-541,Thoennessen2004_RPP67-1187,
Lin2011_SciChinaPMA54S1-73,Pfutzner2012_RMP84-567,Ma2015_PLB743-306}.
Beyond the neutron drip line, there may be the two-neutron radioactivity; 
one example of recent interests is $^{26}$O
\cite{Lunderberg2012_PRL108-142503,Kohley2013_PRL110-152501,
Kondo2016_PRL116-102503}.

It is well known that \emph{clustering effects} are important in atomic nuclei
and cluster structure appears in some stable ones if they are excited to be 
close to some thresholds \cite{Oertzen2006_PR432-43}. 
In exotic nuclei, clustering effects can also emerge in low-lying excited states
and even in ground states. 
For exotic nuclei with much more neutrons, a cluster configuration is 
energetically more favored because it permits a more even distribution of 
valence neutrons as shown in Fig.~\ref{fig:physics}(i) 
\cite{Freer2007_RPP70-2149}.
In one recent experimental study of $^{12}$Be, a $0^+$ resonant state with a
large cluster decay branching ratio was observed \cite{Yang2014_PRL112-162501}.
This observation supports strong clustering effects in $^{12}$Be. 
In addition, in some halo nuclei, 
there appears the so called ``Borromean'' structure, which is shown in 
Fig.~\ref{fig:physics}(j) \cite{Johnson2004_PR389-1}---a bound three-body system
with any two-body subsystems unbound. 

\section{\label{sec:highlights}Highlights of recent progresses}

Next I will highlight some recent progresses on theoretical study of exotic 
nuclear structure. There are indeed many interesting and important works, 
but I can only choose some of them due to the limitation of pages. 
More extensive discussions can be found in Ref.~\cite{Zhou2017_NSC2016}.

\subsection{The weakly bound feature of exotic nuclei}

Many models have been developed to take into account the contribution of 
continua and resonances, see, e.g., Refs.~\cite{Meng2015_JPG42-093101,
Sagawa2015_EPJA51-102,Frederico2012_PPNP67-939,Ji2016_IJMPE25-1641003,
Meng2016_RDFNS-83} 
for recent reviews.
There are many ways to locate single particle resonances. 
Besides the conventional scattering phase shift method 
\cite{Hamamoto2016_PRC93-054328}, 
several bound-state-like approaches \cite{Efros2007_JPG34-R459,
Carbonell2014_PPNP74-55}, such as
the analytical continuation in coupling constant \cite{Tanaka1997_PRC56-562,
Yang2001_CPL18-196,Zhang2004_PRC70-034308,Guo2006_PRC74-024320,
Zhang2012_PRC86-032802,Xu2015_PRC92-024324}, 
the real stabilization method \cite{Zhang2008_PRC77-014312, Zhou2009_JPB42-245001,
Pei2011_PRC84-024311} 
and
the complex scaling method (CSM) \cite{Myo2014_PPNP79-1,
Papadimitriou2015_PRC91-021001R,Shi2014_PRC90-034319,Shi2015_PRC92-054313}, 
are often used to study single particle resonances in atomic nuclei. 
Several other methods, e.g., 
the Jost function method \cite{Lu2012_PRL109-072501,Lu2013_PRC88-024323}, 
the Green's function method \cite{Matsuo2001_NPA696-371,Sun2014_PRC90-054321}, 
the Green's function + CSM \cite{Shi2015_PRC92-054313,Shi2016_PRC94-024302}, 
and solving Schr\"odinger or Dirac equations in the complex momentum 
representation \cite{Li2016_PRL117-062502, Fang2017_PRC95-024311},
have also been implemented in nuclear models.

For describing the contribution from the continua in the mean field level, 
the conventional BCS method suffers from some problems. 
One of them is the non-localization of nucleon density distributions 
\cite{Bulgac1980_nucl-th9907088,Dobaczewski1984_NPA422-103}.
One way to solve partly this problem is to use the resonance BCS (rBCS) approach
\cite{Sandulescu2000_PRC61-061301R,Geng2003_PTP110-921,Zhang2013_EPJA49-77}:
After single particle resonances are located, their contribution can be taken 
into account through the BCS approximation. 
It was also justified that if one, instead of using the conventional BCS 
approximation, makes Bogoliubov transformation and solves 
the Hartree-Fock-Bogoliubov (HFB) equations in $r$ space, the contribution from
continua can be included properly and the nucleus in question is 
localized \cite{Bulgac1980_nucl-th9907088,Dobaczewski1984_NPA422-103}. 
Since then, the HFB models have been developed for spherical
nuclei with continuum either discretized \cite{Dobaczewski1984_NPA422-103,
Dobaczewski1996_PRC53-2809,Yu2003_PRL90-222501,Schunck2008_PRC78-064305} 
or treated with scattering boundary conditions 
\cite{Zhang2011_PRC83-054301,Zhang2012_PRC86-054318,Zhang2017_PRC95-014316}. 
The HFB model was also extended to the study of deformed 
nuclei \cite{Pei2013_PRC87-051302R,Pei2014_PRC90-024317,Nakada2008_NPA808-47}.
In parallel, the relativistic Hartree-Bogoliubov (RHB) or relativistic HFB 
models were established for spherical \cite{Meng1996_PRL77-3963,
Meng1998_NPA635-3,Poschl1997_PRL79-3841,Long2010_PRC81-024308} and deformed 
exotic nuclei including halos \cite{Zhou2010_PRC82-011301R,Li2012_PRC85-024312,
Li2012_CPL29-042101,Chen2012_PRC85-067301}.

\subsection{Deformation effects in nuclear halos}

Based on the deformed RHB model in a Woods-Saxon basis \cite{Zhou2003_PRC68-034323},
shape decoupling effects have been predicted in $^{42,44}$Mg 
\cite{Zhou2010_PRC82-011301R,Li2012_PRC85-024312}: The core of these nuclei is 
prolate, but the halo has an oblate shape.
The generic conditions for the occurrence of halo in deformed nuclei and
shape decoupling effects were given in Ref.~\cite{Zhou2010_PRC82-011301R}.
Later, with a non-relativistic HFB model, Pei et al. predicted that $^{38}$Ne 
has a nearly spherical core, but a prolate halo 
\cite{Pei2013_PRC87-051302R,Pei2014_PRC90-024317}. 
Similar effects has been investigated in Ref.~\cite{Misu1997_NPA614-44} in 
which a square well potential was used and the spin-orbit coupling was neglected. 
These predictions are made for the ground state. It would be interesting to 
study dynamics and excitations of these deformed halo nuclei 
\cite{Fossez2016_PRC93-011305R}.

\subsection{Di-neutron correlations}

Concerning di-neutron correlations, progresses in recent years include the study
of Cooper pairs \cite{Zhang2014_PRC90-034313R}, 
di-neutron correlations \cite{Kobayashi2016_PRC93-024310}, 
di-proton correlations \cite{Oishi2014_PRC90-034303}, 
neutron-proton correlations \cite{Masui2016_PTEP2016-053D01}, and so on. 
For example, the asymptotic form of a neutron Cooper pair penetrating to the 
exterior of the nuclear surface was investigated with the Bogoliubov theory 
in Ref.~\cite{Zhang2014_PRC90-034313R}. 
It was found that Cooper pairs are spatially correlated in the asymptotic large
distance limit, and the penetration length of the pair condensate is universally
governed by the two-neutron separation energy.

There are also lots of theoretical investigations on the soft dipole modes 
\cite{
Paar2007_RPP70-691,
Roca-Maza2012_PRC85-024601,Vretenar2012_PRC85-044317,
Savran2013_PPNP70-210,
Ebata2014_PRC90-024303,
Inakura2014_PRC89-064316,
Papakonstantinou2015_PRC92-034311,
Ma2016_PRC93-014317,
DeGregorio2016_PRC93-044314,
Zheng2016_PRC94-014313,
Nakatsukasa2016_RMP88-045004}. 
In Ref.~\cite{Ebata2014_PRC90-024303}, a systematic study with canonical basis 
time dependent HFB theory reveals a number of characteristic features of 
the low-energy $E1$ modes, e.g., a universal behavior in the low-energy $E1$ 
modes for heavy neutron-rich isotopes, which suggests the emergence of 
decoupled $E1$ peaks beyond $N = 82$.

\subsection{The shell evolution}

It is interesting and instructive to choose the Ca isotope chain as an example 
to discuss the evolution of shell structure and changes of magicity 
because there might be five magic numbers in the Ca isotopes. 
In Ref.~\cite{Wienholtz2013_Nature498-346}, from the measured mass of 
$^{53,54}$Ca, a prominent shell closure at $N=32$ was established. 
This shell closure was later confirmed together with a new one at $N=34$, 
indicated by the fact that the energy of the first $2^+$ state for $^{52,54}$Ca rises 
dramatically \cite{Steppenbeck2013_Nature502-207}. 
Recently the magicity of $N=32$ was shown to persist in Sc isotopes 
\cite{Xu2015_ChinPhysC39-104001}.
The appearance of the shell closures at $N=32$ and 34 have been attributed to 
the evolution of neutron $f_{5/2}$ orbital, which rises due to a weakened
proton-neutron interaction when $Z$ decreases to 20 \cite{Steppenbeck2013_Nature502-207}.

Theoretically there have been many investigations on the magicity of  
$N=32$ and/or 34
\cite{Grasso2014_PRC89-034316,Yueksel2014_PRC89-064322,Wang2015_JPG42-125101,
Li2016_PLB753-97}. 
For example, it has been shown in Ref.~\cite{Grasso2014_PRC89-034316} that the 
like-particle tensor contribution is responsible for these new shell closures
and in Ref.~\cite{Li2016_PLB753-97}, the importance of exchange terms in 
relativistic framework on these shell closures was emphasized. 
However, a recent precise measurement of charge radii in Ca isotopes 
\cite{GarciaRuiz2016_NatPhys12-594} casts some doubts on the magicity at $N=32$. 
If $^{52}$Ca is doubly magic, its charge radius should be smaller than 
that of its neighbors. 
But this is not the case \cite{GarciaRuiz2016_NatPhys12-594}. 
Therefore nuclear magicity in exotic nuclei may be ``local'' 
in the sense that it manifests itself in some nuclear properties but not 
in others, contrast to those traditional magic numbers which are ``global'' or
robust and manifest themselves in ``all'' nuclear properties, e.g., separation
energies, charge radii, $Q$ values of $\alpha$ decays, etc.

\subsection{New radioactivities}

For new radioactivity, without going into details, I'd like to mention that 
there have been some systematic studies with the HFB model 
\cite{Olsen2013_PRL110-222501} and 
predictions were also made with the relativistic continuum 
Hartree-Bogoliubov model \cite{Lim2016_PRC93-014314}.

\subsection{Clustering effects}

There have been many interesting results concerning the theoretical study of 
clustering effects in atomic nuclei. For example, a non-localized or container 
picture was proposed for cluster structure \cite{Zhou2013_PRL110-262501}, 
giant dipole resonances were argued to be a fingerprint of cluster structure 
\cite{He2014_PRL113-032506}, one-dimensional $\alpha$ condensation of 
$\alpha$-linear-chain states in $^{12}$C and $^{16}$O were studied 
\cite{Suhara2014_PRL112-062501} and rod-shaped nuclei were explored at 
extremely high spin and isospin \cite{Zhao2015_PRL115-022501}.

One more thing about clustering effects is that from radius-constrained 
mean field calculations, regardless of non-relativistic 
\cite{Girod2013_PRL111-132503} or relativistic models 
\cite{Ebran2012_Nature487-341,Ebran2014_PRC89-031303R}, one can also obtain the 
cluster structure. 
However, in such kind of studies, one has to take a serious care of 
the truncation of the basis and to ensure a convergence 
\cite{Zhou2015_SINAP-CUSTIPEN}.

\section{\label{sec:summary}Concluding remarks and perspectives}

To summarize, after introducing the following characteristic features and new 
physics connected with exotic nuclear phenomena: the weakly-bound feature, 
the large-spatial extension in halo nuclei, deformation effects in halo nuclei, 
the shell evolution, new radioactivities and clustering effects, 
I have highlighted some recent progresses corresponding to these features.

It should be emphasized that to describe the structure of exotic nuclei, one 
often needs to modify conventional nuclear models or develop new theoretical 
approaches. Nowadays, there are many attempts to unify nuclear models. 
For example, with the fast development of supercomputers, ab initio theories 
can deal with heavier and heavier nuclei, as discussed in Ekstr\"om and Bacca's 
talks. 
Besides that, there are also projects to develop density functional theories 
from first principles \cite{Dobaczewski2016_JPG43-04LT01,Shen2016_CPL33-102103} 
(also mentioned in Liang's plenary talk)
or models based on subnucleon degrees of freedom \cite{Stone2016_PRL116-092501}.

In his talk, Nazarewicz has put atomic nuclei on a table with three pillars. 
This table would not be stable if it had only two pillars, only theory and simulations. 
We need experiments and experimental facilities.
With the development of radioactive ion beam facilities around the world,
including the High Intensity heavy ion Accelerator Facility (HIAF)
in Huizhou, China \cite{Zhou2016_Huizhou-HIAF,Tanihata2016_Huizhou-HIAF} and 
Beijing Isotope-Separation-On-Line Neutron-Rich Beam Facility (BISOL)
\cite{Zeng2015_ChinSciBulletin60-1329}, 
more unstable nuclei would become experimentally accessible,
which will for sure challenge as well as provide opportunities for
theoretical study of exotic nuclear structure.

\section*{Acknowledgements}

Collaborations and/or helpful discussions with
A. Afanasjev,
K. Blaum,
Y. Chen, G. Colo,
L.S. Geng, N.V. Giai,
L.L. Li, H.Z. Liang, W.H. Long, B.N. Lu, H.F. L\"u,
J. Meng,
J. Pei,
P. Ring,
H. Sagawa, J.R. Stone, X.X. Sun,
I. Tanihata,
J. Terasaki, A.W. Thomas, H. Toki,
D. Vretenar, 
N. Wang,
F. R. Xu,
S. Yamaji,
J.Y. Zeng, Y.H. Zhang, S.Q. Zhang, E.G. Zhao, J. Zhao and P.W. Zhao
are gratefully acknowledged.

%


\end{document}